\documentclass[12pt,letterpaper,titlepage, margin=1in]{article}

\usepackage[margin=1in]{geometry} % See geometry.pdf to learn the layout options. There are lots.
\usepackage{graphicx}
\usepackage{amssymb}
\usepackage{latexsym}
\usepackage{amsthm}
\usepackage{amscd}
\usepackage{epstopdf}
\usepackage{amsmath}
\usepackage{caption}
\usepackage{subcaption}
\usepackage{setspace}
\usepackage[utf8]{inputenc}
\usepackage[english]{babel}
\usepackage{csquotes}
\usepackage{float}
\usepackage{tabularx}
\usepackage[square,sort,comma,numbers]{natbib}
\usepackage{url}
\usepackage{hyperref}

\hypersetup{
    colorlinks=false,
    pdfborder={0 0 0},
}

\begin{document}

\begin{titlepage}
    \begin{center}
    \vspace*{2.0cm}
        {\Huge Thesis \par}
        \vskip 5em
        {\large IMF Lending and Economic Growth: An Empirical Analysis of Ukraine \par}
        \vskip 7em
	Submitted by \par
Roman Kononenko \par

 \vskip 3em
In partial fulfillment of the requirements \par
For the Degree of Master of Public Administration \par
With concentration in Economics and Financial Policy \par
Cornell University \par
Ithaca, NY

 \vskip 1em
Advisor: Nancy Chau

  \vskip 3em
\begin{figure}[h!]
\centering
         \includegraphics[width=6cm]{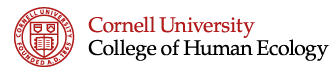}
\end{figure}

Spring 2015

 \end{center}

\end{titlepage}

\newpage

\doublespacing
\tableofcontents
\singlespacing

\newpage

\renewcommand{\thesection}{\Roman{section}}
\section{Abstract}

\doublespacing

\hspace{1cm} This study uses Vector Autoregression (VAR) Methodology as well as Vector Error Correction (VEC) Methodology to examine the existence and direction of causality between economic growth and IMF lending for Ukraine. The paper examines the IMF lending data for the period of 1991-2010. Robust empirical analysis indicates that IMF lending has a negative effect of on Ukraine’s economic growth in the short term. Policy implications of this finding are that, despite short-run decline in economic growth, IMF lending can result in a long-run sustainable growth for Ukraine. For this, policymakers need to ensure that fund’s money are used not only to cover budget’s deficit, but also to finance institutional reforms.

\vspace{5 mm}
\textbf{Keywords: Economic Growth in Ukraine, IMF Lending, Institutional Reforms}

\vspace{5 mm}

\section{Introduction}

\hspace{1cm} After colonial rule of the Russian Empire/Soviet Empire ended and Ukraine regained its independence in 1991, Ukraine needed to rapidly transform its soviet-style plan economy into a market one. This required substantial monetary inflows, since Ukraine itself did not possess all the necessary financial capital. Consequently, starting from 1991 Ukraine has been receiving financial assistance from a number of International Financing Organizations (IFO). As summarized by Ozarina and Alekejeva (2008), the three major IFO borrowers of Ukraine are International Monetary Fund (IMF), International Bank for Restructuring and Development (IBRD) and European Bank for Restructuring and Development (EBRD). By 2006 IMF emerged as the largest lender out of the three IFOs, accounting for 53\% of all funds received\footnote{These percentages were calculated by Ozarina and Alekejeva using the actual funds received by Ukraine, as opposed to the amount decided in the agreement.} by Ukraine from all IFOs. World Bank was second largest lender to Ukraine, and through its subsidiary IBRD, accounted for 38\% of all lending. European Bank for Restructuring and Development had the smallest share of all the IFOs and accounted for only 9\% of all lending. In terms of distribution of IFO funds, the government sector holds the first place as a primary recipient, followed by energy and infrastructure sectors. \par

By the year 2011, cooperation between Ukraine and the IMF has resulted in Ukraine receiving a total of 12.259 billion SPD (approximately 18.28 billion USD) by 2011 (Kurdydyk, 2012). In recent years, Ukraine’s government continued its close cooperation with the IMF and since the 2008 financial crisis has been the recipient of a number of substantial loan packages from the IMF. In particular, a stand-by agreement signed between Ukraine and the IMF in November 2008 has resulted in the IMF approving a 16.5 billion USD loan to Ukraine. Furthermore, in March 2015 the IMF approved a new 17.5 billion USD loan, to be distributed over the course of the next four years, conditional on meeting certain reform criteria. \par

Current scholarship that studies the relationships between the IMF lending and a country’s growth primarily uses a panel data on a large sample of developing countries. Consequently, most papers do not focus on individual states. This paper will be different from the existing papers that analyze effectiveness of IMF programs in that it focuses solely on one developing country, Ukraine. Since IMF has been cooperating with Ukraine starting from 1992, we currently possess nearly 20 years of statistical data that provides us with both a period of almost exclusive decline (1991-2003) and a period of relative\footnote{Ukraine faced a sudden economic slowdown in 2008 as a result of global financial crisis. The reason Ukraine was so vulnerable to it was because its exports largely consisted of only raw materials, mainly metal and metal derivatives, whose prices fluctuated significantly as a result of the 2008 international financial crisis.}  recovery (2003-2010) of Ukraine’s economy. \par

This paper attempts to find out whether IMF had a statistically significant positive or negative effect on Ukraine’s economic growth, without taking into account the degree of IMF conditions’ implementation or the endogeneity of program participation. The Vector Autoregressive (VAR) as well as Vector Error Correction (VEC) Methodologies are used to analyze the data.

\section{Literature Review}

\hspace{1cm} In the 1970s, International Monetary Fund (IMF) switched from lending its funds mainly to developed countries to helping the developing ones achieve macroeconomic stability through financing specific government reforms. In the years following, the fund has been often criticized as not helping the growth of borrowing countries and on the contrary resulting in the reduction of countries’ GDP (e.g. Dreher 2005, Sorokina 2009). George Soros, the billionaire-founder of the Open Society Foundations, an organization that has been instrumental in helping achieve institutional transformation in many Central and Eastern European (CEE) developing countries, has harshly criticized the IMF even before the 2008 financial crisis. He notes that IMF’s interventions have become part of a problem, instead of being part of the solution. As he states in 1999 interview with PBS:
“The institutions, the IMF, is not adequate to meet these circumstances. It adapted itself and did reasonably well in one crisis after another. There was a big international crisis in the '80s ... mainly focused in

\begin{displayquote}
“The institutions, the IMF, is not adequate to meet these circumstances. It adapted itself and did reasonably well in one crisis after another. There was a big international crisis in the '80s ... mainly focused in America. Then you had the Mexican crisis in '94. Now, you have this latest crisis. Here, the IMF method proved to be inadequate. So their intervention became part of the problem, instead of being part of the solution.”
\end{displayquote}

\par
Consequently, a question arises of whether IMF lending has a positive or negative impact on the borrowing countries. The relationship between economic growth and IMF lending has been studied extensively for different countries and time periods. The most recent studies provide mixed answers to this question.  A few of the recent papers found a negative relationship between the economic growth and IMF loans. They concluded that the fund’s original aim of alleviating macroeconomic imbalances in troubled countries and providing long-term growth has failed and in fact resulted in the reduction of output growth. Examples of such papers are Barro and Lee (2005) who used political economy variables as tools to remedy endogeneity issues, and Vreeland (2003) who used counterfactual analysis, with both concluding that IMF loans have negative impact on economic growth. Dicks-Mireaux, Mecagni and Schadler (2000), on the contrary, who likewise used counterfactual analysis, found a positive linkage between IMF loans and output growth. \par

One of the most profound analysis of the effects of the IMF’s lending on country’s growth is a 2005 paper by Dreher. Dreher used panel data from 98 developing countries from 1970-2000 and analyzed how IMF involvement influenced economic growth in program countries. He found that IMF loans have a negative influence on the economy, but it is mitigated if you take government’s compliance with IMF conditionality into account. His conclusion was that indeed, if one takes the degree of government’s implementation of IMF’s conditions into account, then there will be a positive relationship between IMF loans and economic growth. However, despite this, he finds that in the long run the effect is still negative. \par

Another example is a more recent research paper by Binder and Bluhm (2010), who based their research on panel data for 86 countries over the time period from 1975 to 2005. They used random and fixed effects models to capture country-specific effects. Also, they used a two-step maximum likelihood estimator to cope with sample selection issues. Lastly, to condition for government’s degree of program implementation and its institutional features (health, education attained etc.) on economic growth, they used semi-parametric conditional pooling techniques. Their conclusion was that IMF loans’ effectiveness depends on the level of country’s degree of program implementation as well as index of institutional factors (i.e. health, education attained etc.) and that this effect is positive only if the IMF program is implemented to a sufficient degree or if during the program’s participation, a country also achieves improvement in institutional quality.

\section{Model Specification and Analysis of Results}
\subsection{Model}

\hspace{1cm} There are different econometric models one can use to examine the relationship between IMF lending and economic growth of a particular country. Cointegration model is widely used, but such econometric models as Probit, Logit, Tobit, ARIMA, and Vector Autoregression (VAR) could also be used. This paper uses Vector Autoregression (VAR).
The model specification that this paper uses are given below:
\vskip 1em

$nGDP=\alpha_0+\alpha_1 nGDP+\alpha_2 IMF+\epsilon$ \par
$\Delta nGDP_{(t+1)}=\alpha_0+\alpha_1 \Delta nGDP_{(t-1)}+\alpha_2 \Delta IMF_{(t-1)}+\epsilon_{(t-1)}$

Where: nGDP = nominal GDP, in current USD for Ukraine in millions; IMF = the sum of stand-by IMF RCF, FCL loans in current USD in millions. \par
\vskip 1em

Analysis was made using regression techniques that rely on Vector Autoregressive (VAR) Methodology as well as Vector Error Correction (VEC) Methodology to examine the linkage between economic growth and IMF loans. VAR is a dynamic system that treats all variables as endogenous, as opposed to classic assumption that there are exogenous and endogenous variables. In VAR the level of each variable in the system depends on past changes in that variable and all other variables in the system.  A mix of cointegration, error correction and impulse response techniques were used in the regression, whereas the stationarity requirement was checked using the Augmented Dickey-Fuller and Phillip-Perron unit root tests. To correct for non-stationarity, the required number of lags was chosen using Akaike Information Criteria (AIC), Final Prediction Error Criteria (FPEC), Hannan-Quinn Information Criteria (HQIC), Schwarz Information Criteria (SIC), and the Likelihood ratio test.

\subsection{ Data}

\hspace{1cm} The aim of the study is to cover the period since Ukraine regained its independence from the Soviet Union in 1991. However, there a few restrictions that prevent this paper from using the data from the most current year of 2015. Firstly, due to the lack of data for some variables for the most recent years (even though National Bank of Ukraine (NBU) has already provided GDP numbers for 2011-2014, the IMF lending data was not available for the most recent years), consequently the last year that the research analyzes is 2010. Secondly, even though Ukraine began cooperating with the IMF in 1992, it did not get any loans until 1994, and so the first year that this study includes is 1994. Consequently, the total period that this study covers is 17 years (1994-2010).  Most of the macroeconomic variables were collected from the World Bank Database. Furthermore, information from the Ukrainian Statistical Bureau, National Bank of Ukraine (NBU), and Ukraine’s Treasury was used as a source of data, as well as a way check the validity of nominal figures found in World Bank Database.

\subsection{ Time Series Properties of the Variables (Stationarity)}

This part studies the stationarity/non-stationarity of the variables. Both Augmented Dickey Fuller (ADF) and Phillips -Perron (PP) tests were used in order to insure the result. The result of these tests are summarized in Table 1. \par

Our first step in determining stationarity is to plot our variables against time (Figure 1) and observe whether we can visually recognize which variable is non-stationary. It appears that GDP has an upward trend, which indicates that there is a large chance of having non-stationarity in this variable.

\begin{figure}[H]
        \centering
        \includegraphics[width=16cm]{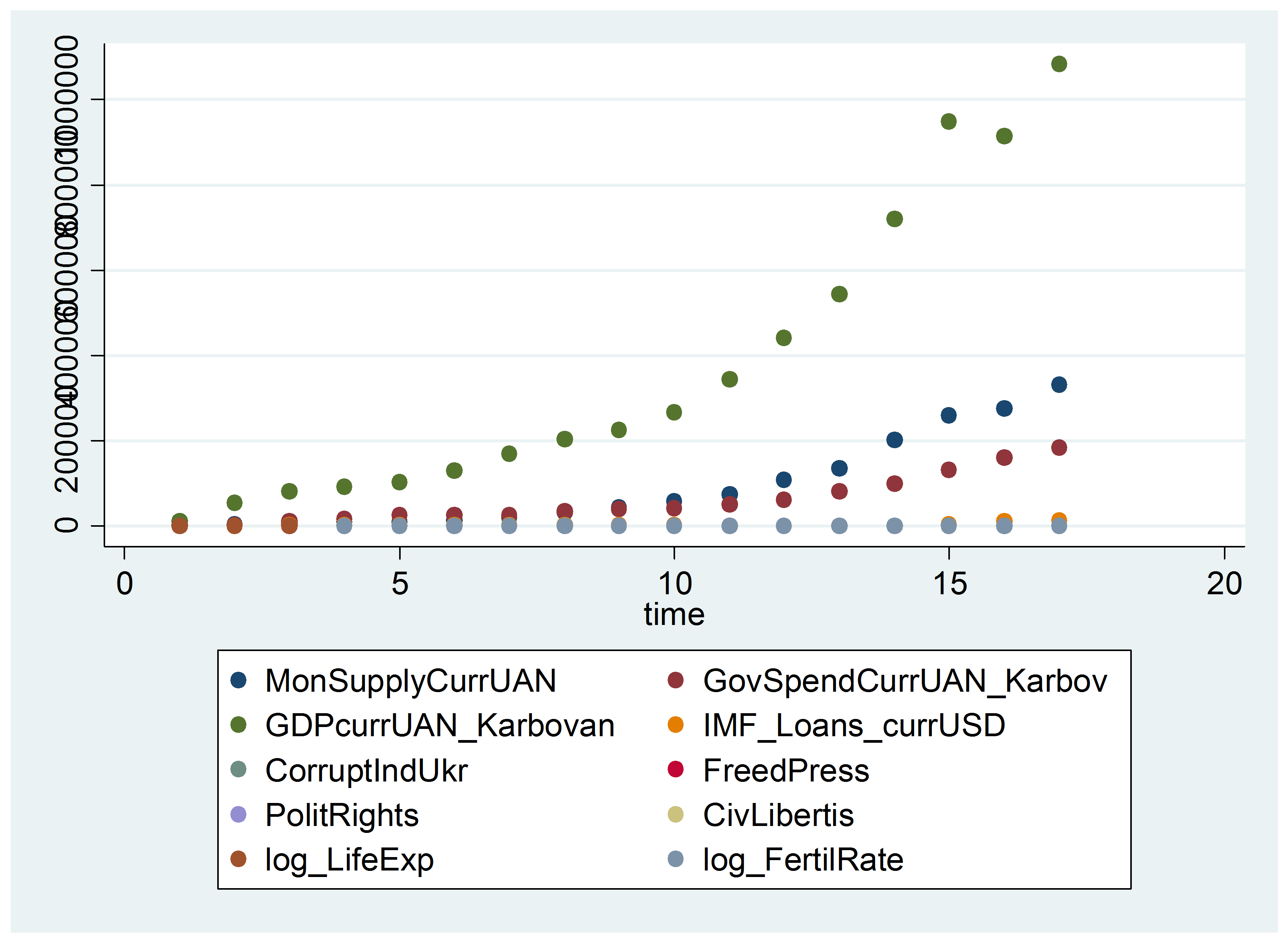}
       % \caption{Stationarity test}
       \caption{}
\end{figure}

\par

\hspace{1cm} Next, after analyzing our variables visually, we want to find out whether there is cross-correlation between Nominal GDP and IMF Loans.  In Figure 2, we see that instead of witnessing a relatively horizontal line, we are witnessing an upward rise and then a slow decrease.

\begin{figure}[H]
        \centering
        \includegraphics[width=16cm]{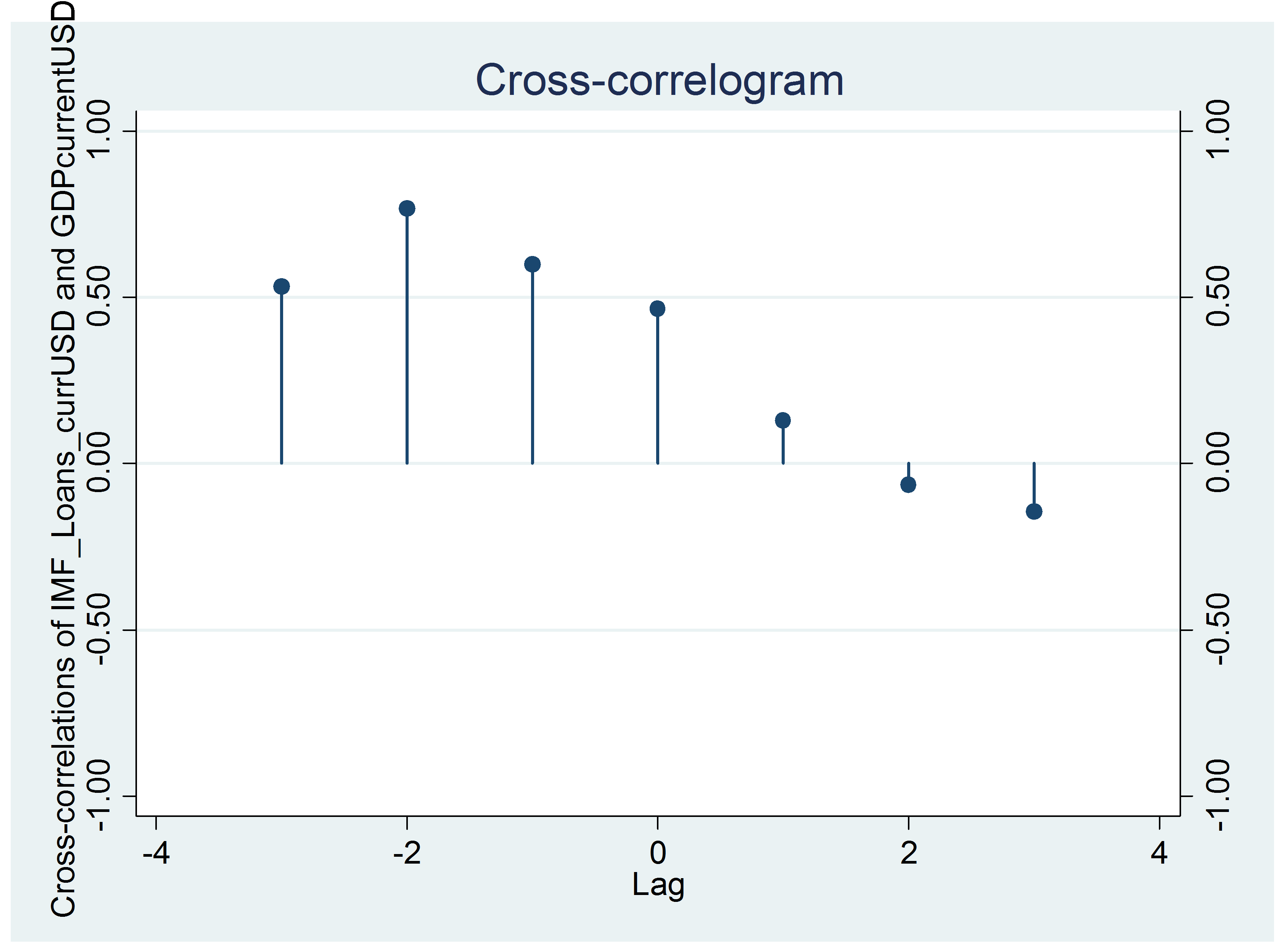}
        % \caption{Auto-correlation  test}
        \caption{}
\end{figure}

\par

Now we check for auto-correlation for each separate variable. Here we present only the graph of nominal GDP (Figure 3) and Gross IMF Loans (Figure 4). As we can see from graph, the IMF loans are stationary, whereas GDP is not.

\begin{figure}[H]
        \centering
        \includegraphics[width=16cm]{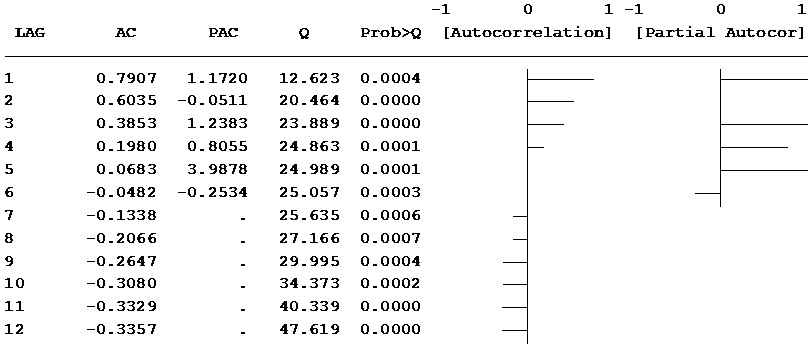}
        \caption{}
\end{figure}

\begin{figure}[H]
        \centering
        \includegraphics[width=16cm]{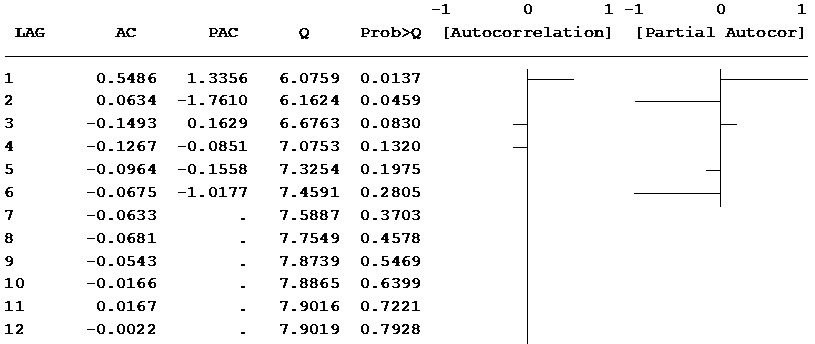}
        \caption{}
\end{figure}

\par

Next, before we proceed to the formal ADF and PP tests, we need to find the optimal lag length for our variables. For this, we use SIC, HQIC or AIC. HQIC suggests that the optimal lag for Nominal GDP in USD is 4 (in conducting ADF, a trend option is added due to the fact that from the plot of GDP on time we noticed an upward trend of GDP that needs to be accounted for). Similarly, using the information criteria, the optimal lag for IMF loan is 1. See Figures 5 and 6 respectively for IMF loans and GDP’s optimal lag lengths.

\begin{figure}[H]
        \centering
        \includegraphics[width=16cm]{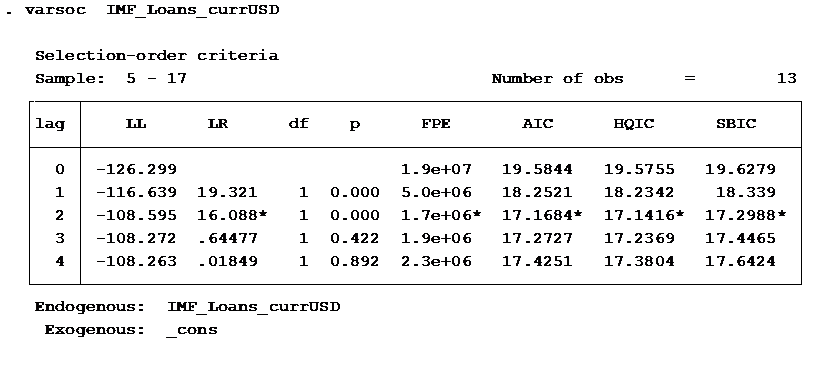}
        \caption{}
\end{figure}

\begin{figure}[H]
        \centering
        \includegraphics[width=16cm]{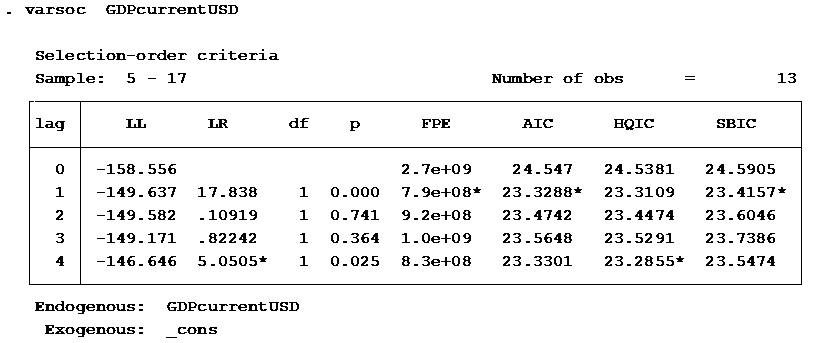}
        \caption{}
\end{figure}

\begin{table}[H]
\centering
\caption{ADF unit root test}
\label{my-label}
\makebox[\linewidth]{
\begin{tabular}{|p{3.8cm}|p{1.4cm}|p{2.5cm}|p{2.5cm}|p{2.5cm}|p{2.5cm}|}
\hline
{\bf Variables}           & {\bf }                 & \multicolumn{4}{l|}{{\bf Augmented Dickey Fuller (ADF) with trend}}                                                    \\ \hline
                          & {\fontsize{8}{7.2}\selectfont Optimal Lags}    & {\fontsize{8}{7.2}\selectfont Level before differencing} & {\fontsize{8}{7.2}\selectfont Level after differencing} & {\fontsize{8}{7.2}\selectfont Initially Stationary} & {\fontsize{8}{7.2}\selectfont Stationary after 1st difference} \\ \hline
{\it GDPcurrentUSD}       & \multicolumn{1}{c|}{4} & \multicolumn{1}{c|}{-1.08}      & \multicolumn{1}{c|}{-4.294}    & \multicolumn{1}{c|}{No}  & \multicolumn{1}{c|}{Yes}              \\ \hline
{\it IMF\_Loans\_currUSD} & \multicolumn{1}{c|}{2} & \multicolumn{1}{c|}{0.0533}     & \multicolumn{1}{c|}{n/a}       & \multicolumn{1}{c|}{Yes} & \multicolumn{1}{c|}{n/a}              \\ \hline

{\it Critical values}     &                        & \multicolumn{4}{l|}{at 10\% =  -3.240}                                                                                              \\ \hline
\end{tabular}
}
\end{table}

\begin{table}[H]
\centering
\caption{PP unit root test}
\label{my-label}
\makebox[\linewidth]{
\begin{tabular}{|p{3.8cm}|p{1.4cm}|p{2.5cm}|p{2.5cm}|p{2.5cm}|p{2.5cm}|}
\hline
{\bf Variables}           & {\bf }                 & \multicolumn{4}{l|}{{\bf Phillips -Perron (PP) with trend}}                                                    \\ \hline
                          & {\fontsize{8}{7.2}\selectfont Optimal Lags}    & {\fontsize{8}{7.2}\selectfont Level before differencing} & {\fontsize{8}{7.2}\selectfont Level after differencing} & {\fontsize{8}{7.2}\selectfont Initially Stationary} & {\fontsize{8}{7.2}\selectfont Stationary after 1st difference} \\ \hline
{\it GDPcurrentUSD}       & \multicolumn{1}{c|}{4} & \multicolumn{1}{c|}{0.6605}      & \multicolumn{1}{c|}{0.0054}    & \multicolumn{1}{c|}{Yes}  & \multicolumn{1}{c|}{I(1)}              \\ \hline
{\it IMF\_Loans\_currUSD} & \multicolumn{1}{c|}{2} & \multicolumn{1}{c|}{}     & \multicolumn{1}{c|}{n/a}       & \multicolumn{1}{c|}{n/a} & \multicolumn{1}{c|}{I(0)}              \\ \hline

{\it Critical values}     &                        & \multicolumn{4}{l|}{at 10\% =  -3.240}                                                                                              \\ \hline
\end{tabular}
}
\end{table}

\subsection {Granger Causality Test}
Now we turn to the Granger Causality test in order to find out whether GDP granger-causes IMF loans, or vice versa or they both interchangeably cause each other. A priori we expect that IMF loans will granger-cause GDP (more loans help stimulate the economy and thus increase GDP), whereas GDP should not granger-cause IMF loans. At first when we run granger causality test, we get counterintuitive result, namely that both GDP and IMF granger-cause each other (Null-hypothesis is that var1 does not granger-cause var2, and we reject it in both cases). See Figure 7

\begin{figure}[H]
        \centering
        \includegraphics[width=16cm]{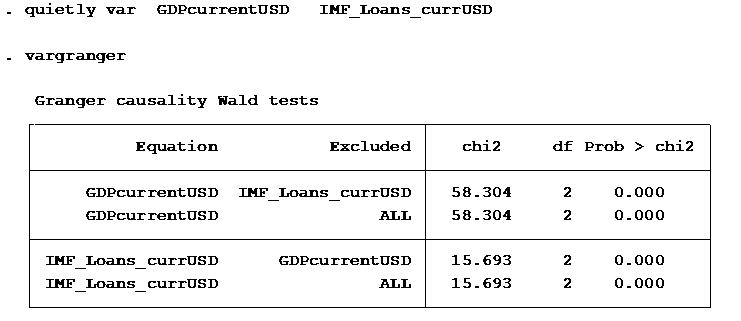}
        \caption{}
\end{figure}

However, this does not take into account the fact that GDP is non-stationary and thus violates the assumption that variance is constant. Consequently, as soon as we fix non-stationarity by differencing GDP by 1, we get the result we expected theoretically, namely that IMF loans granger-cause GDP, whereas GDP does not granger-cause IMF loans. See Figure 8 for details.

\begin{figure}[H]
        \centering
        \includegraphics[width=16cm]{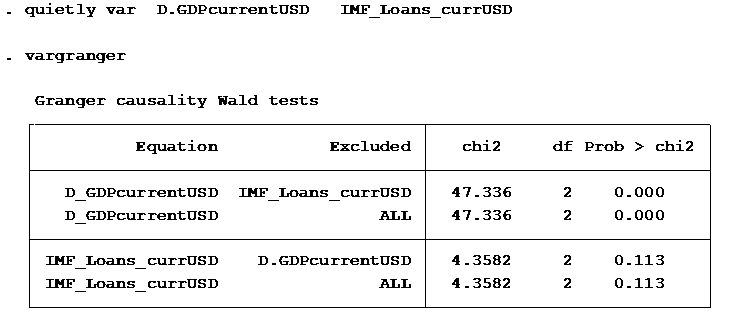}
        \caption{}
\end{figure}

\subsection {Cointegration}
Since we have already discovered both visually and through DAF and PP tests that only GDP is integrated of order one, but IMF-loans are stationary, we no longer need to check for cointegration, since both variables should be non-stationary of order 1.
\subsection {Vector Error Correcting Model}
The Vector Error Correcting Model is capable of taking care of any cointegration between our variables, but since we have already established that we do not have cointegration, the only thing that the VEC model will be taking care of is our variable GDP that is non-stationary. With the VEC, our equation is corrected from long run to short run per period.

\newpage

\begin{figure}[H]
        \centering
        \includegraphics[width=12cm]{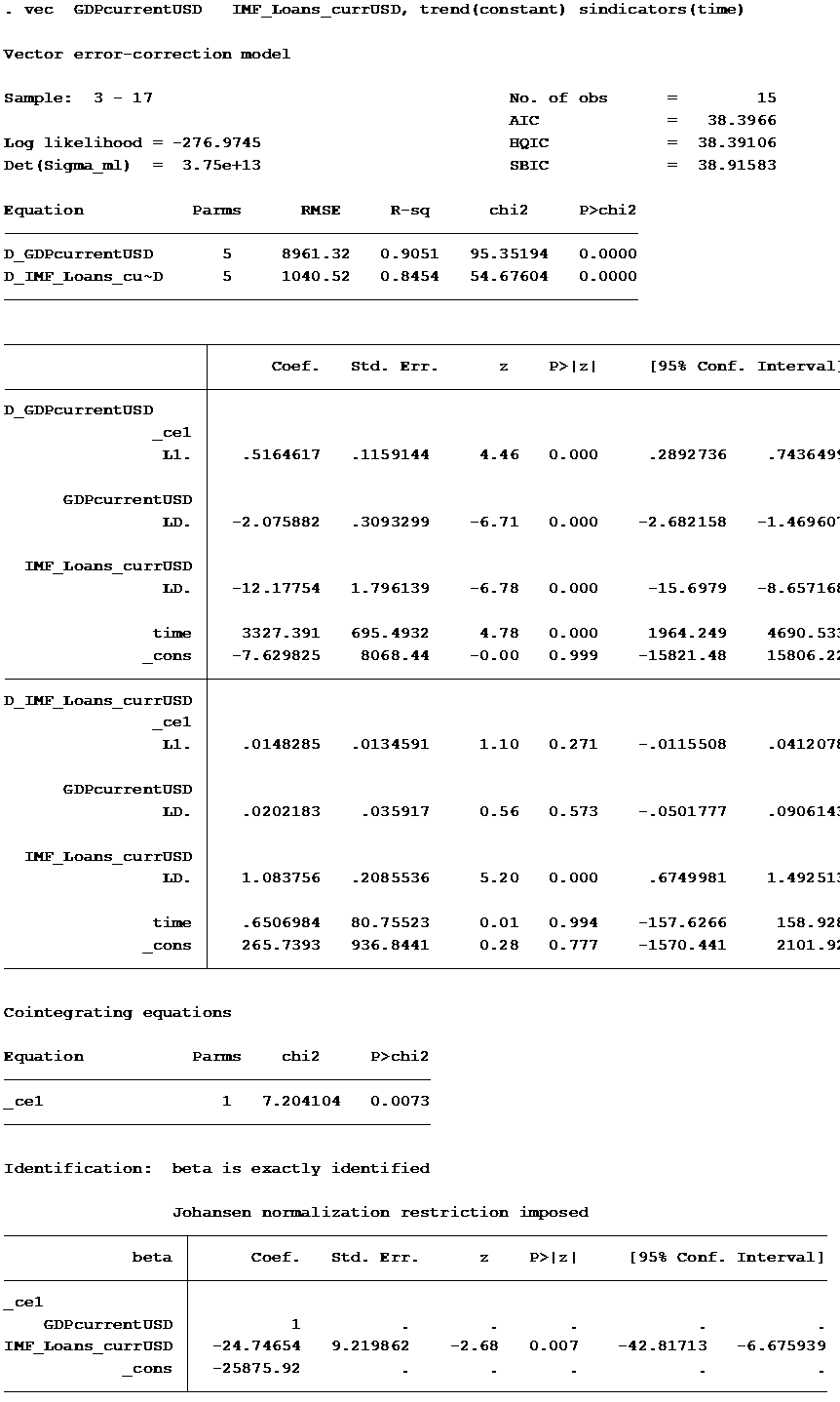}
        \caption{}
\end{figure}

There are a few results worth mentioning. Firstly, as we have noticed, there is a negative coefficient in front of IMF loan, indicating that as IMF loans increase, GDP decreases. Our IMF loan variable is highly significant with a p-value of 0.007.  Initially, we expected there to be a positive relationship between IMF-loans and Ukraine’s economic growth, however, this result is not so surprising if we take into account the fact that we have found the negative effect of IMF lending on Ukraine’s economy only in the short-run. In the long-run, the picture could be quite different, especially if Ukraine’s government successfully completes IMF’s conditions. The overall aim of those conditions is to bring institutional reforms, transparency, and good governance. This will be discussed in more details in the policy implementation section.

\subsection {Impulse Responses}
For the Impulse response function we used up to 4 lags and Figure 10 is what we received.

\begin{figure}[H]
        \centering
        \includegraphics[width=16cm]{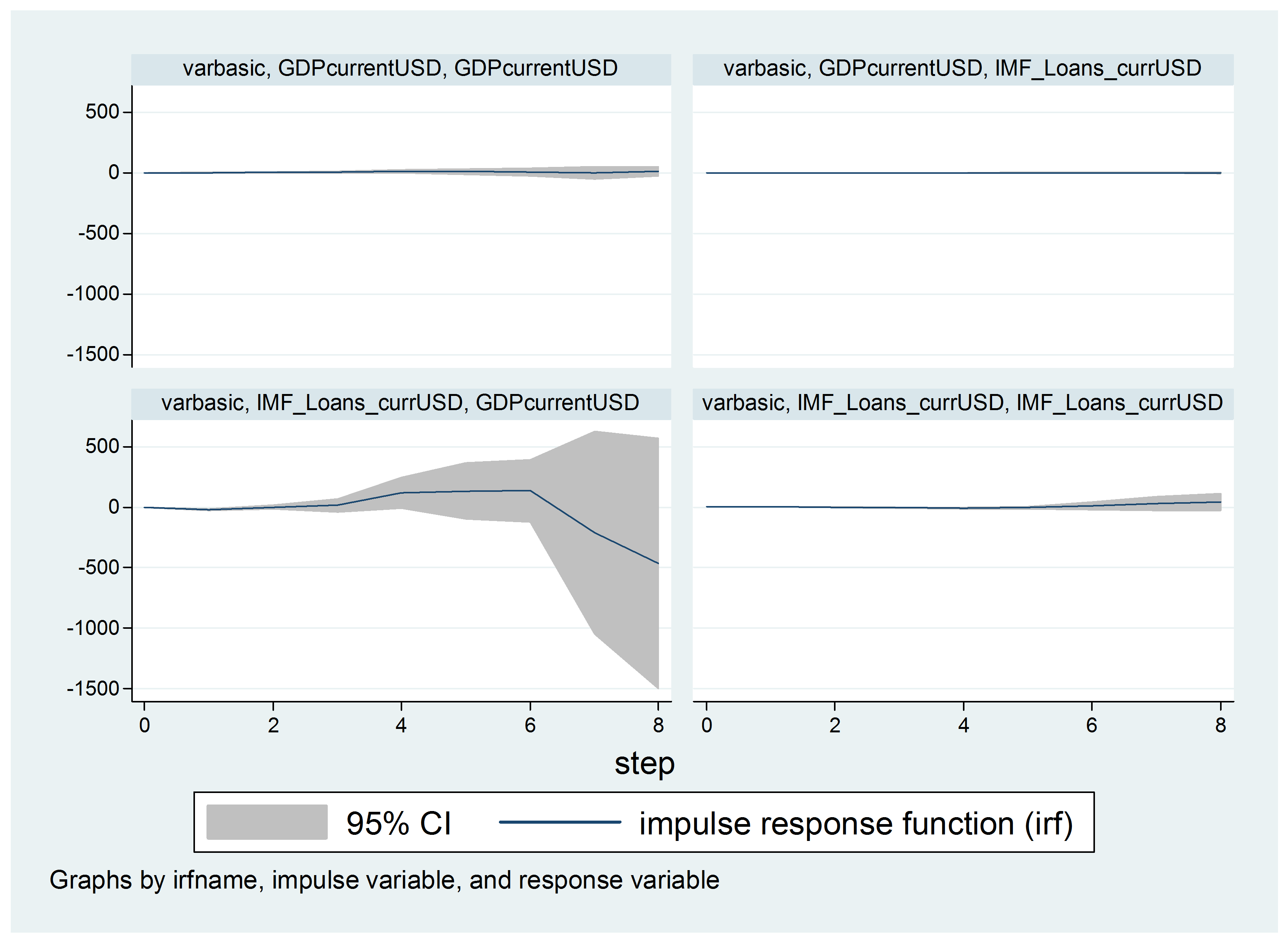}
        \caption{}
\end{figure}

As can be seen from Figure 10, we can observe the largest response when IMF loans have a sudden change. This indicates that Ukraine’s economy is still in a transitional phase and that even after more than 20 years of transitioning from the soviet planned economy into a fully-liberalized market economy, it still depends on the foreign borrowing to stabilize its macroeconomic situation.

\section {Conclusion and Policy Implications}

\hspace{1cm} As this paper has discovered, in the short-run IMF lending has a negative effect on Ukraine’s economic growth. This negative relationship conclusion suggests endogeneity concerns. A number of things could have caused endogeneity, such as omitted variable bias or autocorrelated errors in autoregression among others. A future revision of this paper should address this.

Furthermore, there are other possible reasons for the negative relationship conclusion, such as the need to introduce additional data. As suggested by a few recent papers, in particular Binder and Bluhm (2010), if one accounts for government’s implementation of IMF conditions, then they will uncover a positive relationship between the IMF loans and economic growth at least in the long-run. Nevertheless, this paper did not include the level of fulfilling IMF’s conditions into its regression model, due to a variety of reasons.

One of the primary reasons why the level of implementation of IMF’s conditions was not included was the difficulties in obtaining relevant data. The official database published by IMF, namely Monitoring Fund Arrangements (MONA) does not provide a simple quantitative number that indicates country’s level of IMF programs implementations. Instead, researchers proposed a few methods that speculate how best to transform MONA data into an IMF conditions’ implementation variable. One of the methods proposed suggests calculating the percentage of funds actually drawn in a year, as opposed to the initially decided amount. The problem with this method is that theoretically if all of the IMF’s conditions are not fulfilled by the government, then the country should not get the entire loan amount. As pointed out by Popovs’ka (2009), in practice, however, this has not been the case, and often “Ukraine would chronically not fulfill all of its promised obligations that were originally guaranteed by the heads of Cabinet of Ministers and National Bank of Ukraine, but in the majority of cases the loans were nonetheless given in full.” Another method proposed uses the ratio of the number of conditions actually implemented to the overall number of conditions (and creating a dummy variable of this ratio like, for instance, 1 when the ratio is larger than 50\% and 0 otherwise). There are two major problems with this approach. First, it does not at all compensate for the qualitative difference between the conditions. For instance, in the last couple of years, IMF demanded from Ukraine very serious structural changes to its economy, like, for instance, to increase the retirement age, a change that will affect many future generations much more than if Ukraine fulfilled a condition to have 3\% budget deficit as opposed to 3.5\%. Second, as  Antczak, Markiewicz and Radziwill (2001) pointed out, “unfortunately,  in  countries  of  weak  reform  ownership policies  were  assumed  (and  reluctantly  followed)  just  to  please  the  IMF  and  receive disbursements, rather than to solve the problems of the country.” Consequently, those conditions were mostly done to please the IMF and did not have any real long-term positive effect on the economy, as a result of government’s disinterest in properly implementing them.

One of the reasons as to why the IMF is not more successful in bringing economic growth to Ukraine is because, as stated by Popovs’ka, the IMF provides many suggestions of which macroeconomic indexes need corrections in order to remedy the economy, but very few as to how exactly one is supposed to implement them. Also, as stated by Marynchenko (2010) another reason why IMF loans might not be as effective as they should be is that while in theory all branches of government are supposed to cooperate in order to achieve the highest possible implementation of the IMF’s conditions, in practice they did not. What happened in the past is that executive branch of government, namely the National Bank of Ukraine and Cabinet of Ministers, vouched for IMF loans, whereas the legislative branch, namely Verhovna Rada, instead of backing up those promises with the necessary laws, was torn apart by various political powers that only brought discord in the relationship with the IMF.
Furthermore, another clue as to why the IMF is not more successful in Ukraine, could be found in Podvigin (2010) and Antonenko (2009) works, who indicate that a significant portion of IMF loans has been historically used to repay previous debts and government deficit. Most economist would agree that large government loans are only justified if they are invested in projects that will have a long term positive effect on the whole economy. For example, investment into education, health, infrastructure projects (such as bridges, roads, or airports), or long-term energy diversification investments that would help Ukraine achieve better energy security. When loans are spent on repaying previous loans and covering the budget deficit, this becomes the problem for the future generations that will have to face the initial loan plus interest that might prove to be an impossible burden considering that their predecessors did not stimulate economy through significant projects that would help increase GDP growth.

Consequently, the main policy implication of this finding for Ukraine’s government is that they should accept the fact that in the short-run IMF lending will have an almost uniformly negative impact on Ukraine’s economic growth. However, those policymakers should also realize that the economic slowdown will be only short-term and if they manage to find political consensus and fully implement all the necessary structural reforms proposed by the IMF, Ukraine can achieve a long-term sustainable growth. The key driving force for this sustainable growth is full implementation of IMF’s conditions, which will help achieve a fully-transformed and properly functioning institutions, transparency, and good governance.

\newpage

\bibliography{mybib}{}
\bibliographystyle{amsplain}
\nocite{*}

\newpage
\section {Appendix}

\begin{figure}[H]
        \centering
        \includegraphics[width=16cm]{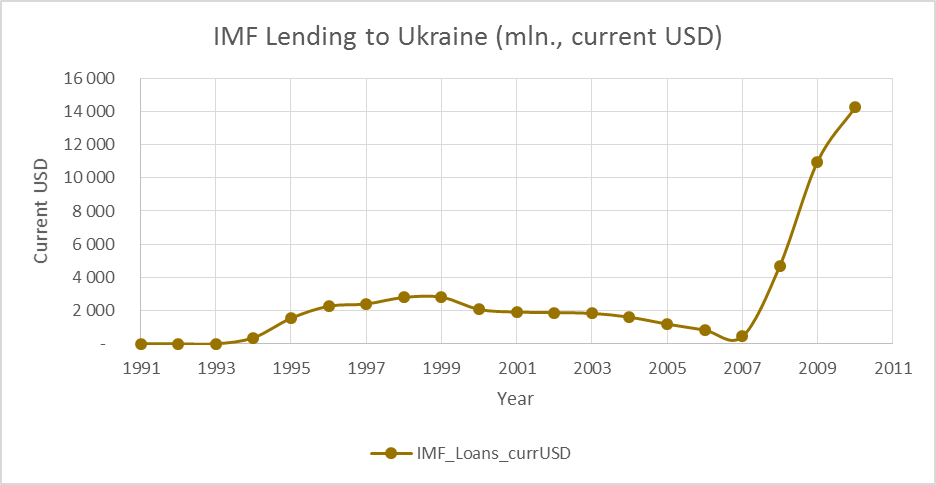}
        \caption{Exhibit 1. IMF Lending to Ukraine in millions USD}
       
\end{figure}

\begin{figure}[H]
        \centering
        \includegraphics[width=16cm]{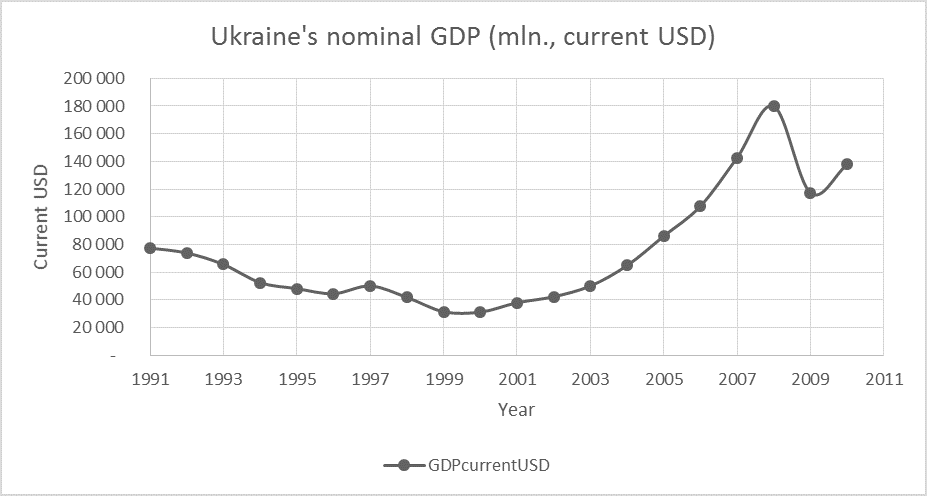}
        \caption{Exhibit 2. Ukraine’s nominal GDP in millions USD}
       
\end{figure}

\end{document}